\documentclass[aoas,nameyear,seceqn,rotating,noautosecdot,dvips]{arximspdf}
\usepackage{graphicx}

\doi{10.1214/10-AOAS348}
\volume{4}
\issue{4}
\pubyear{2010}
\firstpage{1976}
\lastpage{1999}

\makeatletter
\newcommand{\eqref}[1]{(\ref{#1})}
\renewcommand{\citep}[1]{(\citeauthor{#1} \citeyear{#1})}
\newproclaim{remark}{Remark}
\newtheorem{thmm}{Theorem}[section]
\newtheorem{lemm}{Lemma}[section]
\makeatother

\begin{document}
\begin{frontmatter}

\title{An imputation-based approach for parameter estimation in the
presence of ambiguous censoring with application in industrial
supply chain}
\runtitle{Parameter estimation in reliability}

\begin{aug}
\author[a]{\fnms{Samiran} \snm{Ghosh}\ead[label=e1]{samiran@math.iupui.edu}\corref{}}
\runauthor{S. Ghosh}
\affiliation{Indiana University--Purdue University}
\address[a]{Department of Mathematical Sciences\\
Indiana University--Purdue University\\
Indianapolis, Indiana 46202-3216\\
USA\\
\printead{e1}} 
\end{aug}

\received{\smonth{5} \syear{2008}}
\revised{\smonth{3} \syear{2010}}

%
\begin{abstract}
This paper describes a novel approach based on ``proportional
imputation'' when identical units produced in a batch have random but
independent installation and failure times. The current problem is
motivated by a real life industrial production--delivery supply chain
where identical units are shipped after production to a third party
warehouse and then sold at a future date for possible installation.
Due to practical limitations, at any given time point, the exact
installation as well as the failure times are known for only those
units which have failed within that time frame after the
installation. Hence, in-house reliability engineers are presented
with a very limited, as well as partial, data to estimate different
model parameters related to installation and failure distributions.
In reality, other units in the batch are generally not utilized due
to lack of proper statistical methodology, leading to gross
misspecification. In this paper we have introduced a likelihood
based parametric and computationally efficient solution to overcome
this problem.
\end{abstract}

%

\end{frontmatter}
%
\section{\texorpdfstring{Introduction: Background of the problem.}{Introduction: Background of the problem}}
After the production process, consumer goods are often distributed
through multi-step channels, giving rise to the term
``production--delivery'' supply chain. An exception to this practice
is ``just-in-time'' manufacturing where a product is assembled and
shipped directly only upon the request of a customer, which is quite
popular in the personal computer industry. However, for most
consumer products, items produced by a company are not shipped
directly to the final customer. The traditional route for any large
scale industrial operation is to ship the manufactured products to a
warehouse. The warehouses are often maintained by third party
retailer/shops, from where the products are sold and installed at a
future date to the final customer. Due to geographic as well as
company--retailer relationship, once the batch is shipped, it is
often unknown to the producing company whether a specific unit is
working or is still not installed, until and unless the unit stops
working and the final customer claims a warranty at a future date.
At that point in time the data on the failed unit becomes
``complete'' in a sense that we know exactly its installation as well
as failure time. For all other units it is not known (hence
``partial'' information only) whether they are working or are not at
all installed. The above setup is quite common in practice in many
industrial supply chains, giving rise to a situation where in-house
engineers face a dilemma regarding the optimal usage of available
information. The untimely failure of a unit is always costly
to the producer from the warranty perspective [\citet{alber}]. Also,
after infant mortality, reliability assessment and future lifetime
prediction at an early stage of the product lifespan is advantageous
for appropriate customer satisfaction issues.

Reliability estimation requires knowledge of the population at risk
and the reliability of each unit of the population. The major
objective is always to acquire timely information of interest on
failure modes. However, in the presence of both ``complete'' and
``partial'' information, current practice is to estimate relevant
reliability information by using those units which have completed
their life cycle (i.e., ``complete'' portion only), while not
utilizing the ``partial'' information [\citet{alber}; \citet{kece}]. The
primary reason for this is the absence of any established
methodology for dealing with the current situation. This clearly
makes the inferential procedure suboptimal. In this article we
adopt a proportional imputation based approach to yield a practical
solution to the situation described above. The thrust of this paper
is the estimation of the unknown parameters under the assumption
that we know the actual parametric distribution of installation as
well as failure time. The more general problem of unknown
distributional form for either installation or failure time (or
both) is not considered here and is left for future work.

The rest of the article is organized as follows. In the first three
sections we present notation and a theoretical justification of the
proposed methodology. Section \ref{chl} presents the algorithm for
proportional imputation. The connection between the exact likelihood
based approach and our proposed algorithm is described in Section \ref{sec6}.
Section \ref{sec7} describes the simulation performance of our algorithm. We
also include the analysis of industrial furnace data in Section \ref{motiv}.
We conclude the article with some discussion.

\section{\texorpdfstring{Notation and mathematical setting.}{Notation and mathematical setting}}\label{sec2}
The problem of interest is motivated from a large industrial company
producing residential furnace components. The units are produced and
shipped within the continental USA via multiple channels. However,
the general description of the problem and our solution is neither
dependent on a specific company nor confined to a specific
commodity. Rather, our proposed solution will have a broader
application since the setup is common to many production delivery
supply chains. Consider a setup in which $N$ identical units are
produced in a batch, which are then shipped to a warehouse. These
units will be installed only after being purchased by the customer
at some future date. We assume there exists no substantial time lag
between purchase and actual installation of unit/units. Purchase and
installation will be considered as the event of interest, and the
time in which this transpires will be referred to as the
``installation time.'' Consider a fixed end of study time $T_0$. The
general data description at hand is rather simple. For a particular
unit we either know both the installation and failure times or know
nothing at all. In fact, for many units at time $T_0$, their current
status will be unknown due to the fact that they have not yet failed
either due to noninstallation or are still in working condition.
Let $X$ $(\sim F_X(\cdot))$ and $T$ $(\sim F_T(\cdot))$ denote the
continuous random variables corresponding to installation time and
failure time and which are assumed to be independent of each other.
In this paper we assume that $F_X(\cdot)$ and $F_T(\cdot)$ are completely
specified but with unknown parameters. We denote the random set
$\Omega=\{i\in\{1,2,\ldots,N \}\dvtx  X_i+T_i\leq T_0\}$ to be the set
of indices of the completely observed units. Let $C$ denote the
cardinality of $\Omega\dvtx C=|\Omega|=\sum_{i=1}^N I\{X_i+T_i \leq
T_0\}$. Following standard results in survival/reliability analysis,
the complete likelihood for the above setup is
%
%
\begin{eqnarray}\label{e2}
\hspace*{8.5pt}\qquad L(F_X, F_T) &=& \prod_{i=1}^N [ f_{X,T} (x_i,t_i) I\{x_i+t_i
\leq T_0\} ]^{\tau_i} [P\{X+T>T_0 \}
]^{1-\tau_i}
\nonumber
\\[-8pt]
\\[-8pt]
\nonumber
&\propto& \biggl\{ \prod_{i\in\Omega} f_{X,T} (x_i, t_i)\biggr\}
\biggl\{S_X (T_0) + \int_0^{T_0} S_T (T_0-x)\,dF_X
(x)\biggr\}^{N-C},\!\!\!\!\!\!\!\!\!
\end{eqnarray}
where $\tau_i$ is an indicator of whether the $i$th unit is observed
or not for $i=1,2,\ldots,N$. The above likelihood is difficult to
maximize numerically except for the very restrictive case when $X$
and $T$ are independent and identically distributed (i.i.d.)
according to an exponential distribution. For the other popular
reliability distributions (e.g., Weibull, Gamma), the above
likelihood is difficult to maximize due to excessive flatness,
especially when $C\ll N$. In the furnace data described in Section
\ref{motiv}
and also in other simulation studies, the $\frac{C}{N}$ ratio is on
average $40\%$ or below. With only this much data the above
likelihood essentially becomes very flat and brute force
optimization often produces unstable estimates with large variances.
For more details on this see the simulation studies in Section~\ref{sec7}.
Next we provide a proportional imputation scheme that has close
connection with the above likelihood, yet it employs a search
strategy parallel to Monte-Carlo-based approaches which is
computationally faster and produces stable estimates.

\subsection{\texorpdfstring{Standard practice and an alternative formulation.}
{Standard practice and an alternative formulation}}
For notational simplicity and without loss of generality, we assume
that the first $C$ units are observed or, in other words, we have
complete information for $\{x_i, t_i\}_{i=1}^C$. Notably, the
manufacturer knows nothing about a unit under two circumstances.
First, if $X>T_0$, that is, the unit is not being installed until time
$T_0$ and denoted as event $B$. Second, $X<T_0$ but $T>T_0 -X$, that is,
the unit is installed but still in operation and denoted as event
$D$. Since exact likelihood is difficult to use, traditional
practice is of two forms [\citet{alber}; \citet{kece}]. The most simplistic
approach is to think that only $C$ units are produced. Since we will
have complete information for all of them, we may use standard
theory to estimate model parameters corresponding to $X$ and $T$
under specific distributional choices. The other practice is to
think that we have $C$ units not from the full distribution but
rather from the truncated distribution of both $X$ and $T$ (i.e.,
observed if $X<T_0$ and $T<T_0$). Then under some specific
distributional assumptions (popular choices are Exponential, Weibull,
etc.) the MLE or rank egression based approaches are used for
parameter estimation [\citet{wang}; \citet{john}; \citet{michael}]. Both of these
approaches will produces erroneous estimates for the setup
considered. The situation will be much simpler if it is also known
for a specific ``noninformative'' unit whether it is under the event
$B$ or $D$. This knowledge, if available, will enable us to render
the case as Type-1 right censoring at $T_0$ either on $X$ (under
$B$) or on $T$ (under $D$) and then follow the usual theory of
estimation with censored data [\citet{meek}; \citet{klein}]. Unfortunately,
practical considerations suggest that even this information will not
be available under most producer--retailer setups resulting in
``ambiguous'' censoring. This is unavoidable unless the producer
company has an agreement with the retailer to get in-time unit
specific sales information. This involves monetary implications and
often short-term cost cutting actions get higher priority.

In this article we took an alternative route to impute the
installation time ($X$) for those units under $D$, that is, installed
but not failed. Note that if we know or can successfully impute the
installation time and assume that the unit is still working, this
essentially means the failure time is being censored. This enables
us to use standard methodology to estimate the model parameters [see
\citet{meek}; \citet{klein}]. However, the crucial question is not only how
to impute the unobserved installation time, but also how many units
are needed to be imputed. Next we present the theory of an
interesting computational approach to achieve this task based on a
proportional sampling imputation scheme.

\section{How many to sample and where to sample from?}\label{sec3}
 In the
parametric setup we generally assume some distributional form for
$X$ and $T$, Weibull and Exponential being the most popular choice
to reliability engineers [\citet{alber}]. Our present methodology is
general in the sense that it does not depend on any specific
distributional choice for both $X$ and $T$. Note that for $C$
complete units we have samples from three conditional distributions,
namely:
\begin{enumerate}[]
\item$x|X+T\leq T_0$;
\item$t|X+T\leq T_0$;
\item$x+t|X+T\leq T_0$.
\end{enumerate}
It is not difficult to formalize an estimation procedure if we have
samples from $\{x|X\leq T_0\}$. However, the identity
\[
f_X (x|X+T\leq T_0) = \frac{f_X (x|X\leq T_0)F_T (T_0 -x)F_X(T_0)}{F_{T+X}
(T_0)}
\]
implies
%
%
\begin{eqnarray}
f_X (x|X\leq T_0) &=& \frac{f_X (x|X+T\leq T_0)F_{T+X}
(T_0)}{F_T (T_0 -x)F_X(T_0)}\nonumber\\
&\propto& f_X (x|X+T\leq T_0)F_T^{-1} (T_0 -x)\\
&\propto& f_X (x|X+T\leq T_0)\{1-S_T (T_0 -x)\}^{-1}\nonumber.
\end{eqnarray}
\begin{remark*} Note that the number of samples (if available) from
$\{x|X\leq T_0 \}$ will be larger than that from $\{x|X+T\leq
T_0\}$. Hence, we have the identity, $\#\mbox{ samples }\{x|X\leq T_0
\}- \#\mbox{ samples }\{x|X+T\leq T_0 \}=\#\mbox{ samples }\{x|X\leq
T_0 \cap T>T_0 - X\}$. We will try to impute this difference (or
unobserved installations) via proportional sampling.
\end{remark*}

The above calculation shows why the assumption that the samples are
from right truncated and independent distributions is not valid.
Even though $X$ and $T$ are assumed to be independent, the very
nature of the ``installation-failure'' setup will make them
intrinsically dependent. Hence, it will be wrong to carry out
separate estimation of the parameters of the distributions of $X$
and $T$ under the truncation assumption, as in reality we do not
have samples from $\{x|X\leq T_0\}$ and $\{t|T\leq T_0\}$. Next we
have exploited this mutual dependence of $X$ and $T$ via a sampling
and imputation based approach.

\subsection{\texorpdfstring{Proportional imputation scheme.}{Proportional imputation scheme}}\label{sec3.1}
To estimate the number of imputations necessary, let us denote the
random variable $V=\sum_{j=1}^N V_j$, where
\[
V_j=\cases{
1, & \quad $\mbox{if $j${th} unit is installed on or before } T_0,$
\vspace*{2pt}\cr
0, & \quad  $\mbox{otherwise.}$}
\]
Hence, $P[V_j=1]=P[X\leq T_0]=F_X (T_0)$ and $V_j \sim \operatorname{Bernoulli}(F_X
(T_0))$.\break Under the assumption that units are identical and
independent, $V \sim \operatorname{Binomial}(N, \break F_X (T_0))$. Hence, $E[V]=NF_X
(T_0)$ and since $C$ units are already observed, we need to impute
for $NF_X (T_0)-C$ units. Of course, $NF_X (T_0)-C$ need not be an
integer and so we round it up to produce a sensible estimate. We use
$[\cdot]$ notation to denote this rounding procedure. All these make
sense provided we know the parameters in $F_X(\cdot)$, but, in fact, the
main purpose of this paper is to estimate those parameters. However,
for the time being let us assume that some crude estimates of these
parameters are available. We will describe exactly how to get such
accurate estimates in Section \ref{chl}.

\begin{figure}

\includegraphics{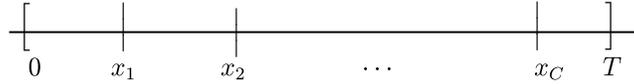}

\caption{Schematic diagram of $C$ observed installations.}
\label{fig1}
\end{figure}

Without loss of generality, we assume $C$ units are ordered in the
sense that $x_i<x_{i+1}$ for $i=1,\dots,C-1$. The observed
installations are depicted in Figure~\ref{fig1}.
These installations produce a natural $C+1$ partitioning of the
study interval, that is, $[0, T_0]$. Due to the continuous
distributional choice for $X$, we consider the case with no ties.
However, we remark that the case with ties can be handled with minor
modifications. The probability of a unit being installed in the
interval $[x_k, x_{k+1}]$ is given by
$P[x_k<X<x_{k+1}]=F_X(x_{k+1})- F_X(x_k)$. An installed unit will
remain unobserved if it does not fail by $T_0$. So the conditional
probability of remaining unobserved is given by
%
%
\begin{equation}\label{a1}
P[T>T_0 - X|x_k<X<x_{k+1}]=\frac{ \int_{x_k}^{x_{k+1}} S_T (T_0 -x)f_X
(x)\,dx }{F_X(x_{k+1})-
F_X(x_k)}.
\end{equation}
Next we present a theorem for the above conditional probability if
the interval $[x_k, x_{k+1}]$ becomes narrower, that is, $x_{k+1}
\downarrow x_k$.
\begin{thmm}\label{t1}
$\lim_{x_{k+1} \downarrow x_k } \frac{ \int_{x_k}^{x_{k+1}} S_T (T_0
-x)f_X (x)\,dx }{F_X(x_{k+1})-
F_X(x_k)}=S_T (T_0 - x_k)$, provided $f_X{(x_{k+1})}\neq0$.
\end{thmm}
\begin{pf}
This follows by application of l'Hospital's rule.
\end{pf}
\begin{remark*} This indicates that if $x_{k+1} \downarrow x_k$, then
the probability of survival (i.e., remaining unobserved) for a unit
installed exactly at $x_k$ will be $S_T (T_0 - x_k)$.

Now using equation (\ref{a1}), the joint probability of a unit being
installed in $[x_k, x_{k+1}]$ and then remaining unobserved is
%
%
\begin{equation}\label{a2}
\qquad P[(x_k<X<x_{k+1}) \cap(T>T_0 - X)]= \int_{x_k}^{x_{k+1}}
S_T (T_0 -x)f_X (x)\,dx.
\end{equation}
Due to the nonincreasing property of the survival function, it is
easy to see that
%
%
\begin{eqnarray}\label{a3}
S_T (T_0 -x_k) \int_{x_k}^{x_{k+1}} f_X (x)\,dx &\leq&\int
_{x_k}^{x_{k+1}} S_T (T_0 -x)f_X
(x)\,dx
\nonumber
\\[-8pt]
\\[-8pt]
\nonumber
&\leq& S_T (T_0 -x_{k+1}) \int_{x_k}^{x_{k+1}} f_X (x)\,dx.
\end{eqnarray}
We would like to use the above inequality to approximate equation
(\ref{a2}) via
%
%
\begin{eqnarray}\label{a4}
I_{k+1}&=& P[(x_k<X<x_{k+1}) \cap(T>T_0 - X)]
\nonumber
\\[-8pt]
\\[-8pt]
\nonumber
&\simeq& \frac{S_T (T_0 -x_k)+S_T (T_0 -x_{k+1})}{2}[ F_X (x_{k+1})-
F_X(x_k)].
\end{eqnarray}
\end{remark*}
\begin{remark*} Note if $T_0\downarrow$ but $[x_k, x_{k+1}]$ remains
fixed with $x_{k+1}\leq T_0$, then $I_{k+1}\uparrow$ due to the
monotone decreasing property of the survival function. Conversely,
if $T_0\uparrow$, then $I_{k+1}\downarrow$. The approximation for
$I_{k+1}$ given in equation (\ref{a4}) works very well provided the
observed installation times are not very sparse over $[0, T_0]$.
Next, we present a theorem characterizing unobserved installation
times over different regions.
\end{remark*}
\begin{thmm}\label{t2}
Let $x_k\in(x_{k-1},x_{k+1})$. Then $P[T>T_0 -
X|x_{k-1}<X<x_{k}]\leq P[T>T_0 - X|x_k<X<x_{k+1}]$.
\end{thmm}

The proof is provided in the \hyperref[app]{Appendix}. Theorem \ref{t2} implies that
the probability of remaining
unobserved increases as the installation time gets closer to the end
of study time $T_0$. Equation (\ref{a4}) characterizes the probability
of a single unit being installed in $[x_k, x_{k+1}]$ but remains
unobserved until $T_0$. Note that we have $C+1$ such intervals in
$[0,T_0]$. Hence, the expected number of unobserved installations in
$[x_k, x_{k+1}]$ is
\[
\alpha_{k+1}=\frac{\{NF_X
(T_0)-C\}I_{k+1}}{\sum_{j=0}^C I_{j+1}},
\]
with the identity
$\sum_{k=0}^C\alpha_{k+1}=NF_X (T_0)-C$.
\begin{lemm}\label{l1}
$\sum_{k=0}^C I_{k+1}= \sum_{k=1}^C \frac{F_X (x_k)}{2} [ S_T
(T_0 - x_{j-1})- S_T (T_0 - x_{j+1}) ]+ F_X (T_0)\frac{1+ S_T
(T_0 - x_c)}{2}$, where $x_0=0$ and $x_{C+1}=T_0$.
\end{lemm}
\begin{pf}
Note that $I_{k+1}= \frac{S_T (T_0 -x_k)+S_T (T_0
-x_{k+1})}{2}[ F_X (x_{k+1})- F_X(x_k)]$. Hence,
\begin{eqnarray*} 
\sum_{k=0}^C I_{k+1} &=& \sum_{k=0}^C \frac{S_T (T_0 -x_k)+S_T (T_0
-x_{k+1})}{2}[
F_X (x_{k+1})- F_X(x_k)]\\
&=& [F_X (x_{1})- F_X(x_0)]\frac{S_T (T_0 -x_0)+S_T (T_0
-x_{1})}{2}\\
& &{} + [F_X (x_{2})- F_X(x_1)]\frac{S_T (T_0 -x_1)+S_T (T_0
-x_{2})}{2}\\
& &\hspace*{5pt}{} \vdots\\
& &{} + [F_X (x_{C+1})- F_X(x_C)]\frac{S_T (T_0 -x_C)+S_T (T_0
-x_{C+1})}{2}.
\end{eqnarray*}
After cancelling successive terms and setting $S_T(0)=1$, we
complete the proof.
\end{pf}

Note that even if the distributional forms for $X$ and $T$ are
known, $\alpha_{k+1}$ will still not be available if we do not know
the parameters of $F_X(\cdot)$ and $F_T(\cdot)$. In Section \ref{chl} we
will propose a general iterative approach for estimating these
parameters which in turn will yield the estimate
$\widehat\alpha_{k+1}$ for $k=0,\ldots,C$. In practice, we use
$[\widehat\alpha_{k+1}]$ for obvious reasons. We would like to put
forward a sampling based approach to impute these unobserved
installation times in Section \ref{chl}. We denote the random set
$\Gamma=\{i\in\{1,2,\ldots,N \}\dvtx (X_i\leq T_0)\cap(X_i+T_i >
T_0)\}$ with $|\Gamma|=\sum_{k=0}^C[\widehat\alpha_{k+1}]$ being the
number of imputed samples of $X$. In this situation, by combining the
observed and imputed samples we have the case of type-1 right
censoring for the installation time $X$. The likelihood for $X$ is
then given by
%
%
\begin{equation}\label{a51}
L_X= \biggl\{\prod_{i\in\Omega\cup\Gamma} f_X (x_i)\biggr\}
S_X(T_0)^{N-C-|\Gamma|} ,
\end{equation}
which we need to maximize with respect to the parameters to obtain
the ML estimates.

\begin{figure}

\includegraphics{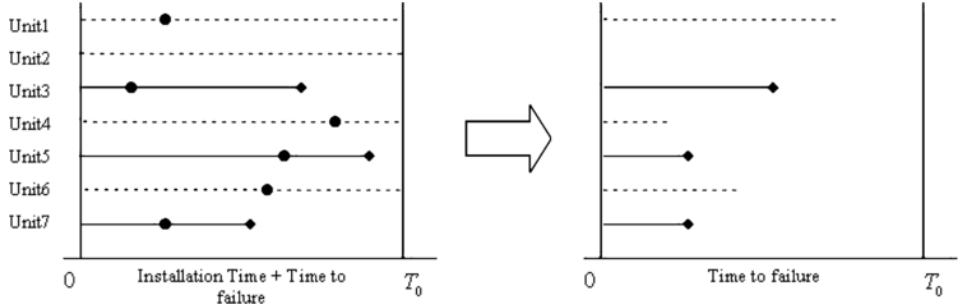}

\caption{Schematic diagram (on left) until observation time $T_0$
with $N=7$ and $C=3$. A ``$\bullet$'' indicates an installation and a
``$\blacklozenge$'' indicates a failure. A solid line indicates an
observed unit (i.e., $X+T\leq T_0$). A dashed line indicates an
unobserved unit [i.e., either $\{X> T_0\}$ or $\{(X<T_0) \cap(T>T_0
-X)\}$]. Note that units 1, 4 and 6 are installed but still working, while
unit 2 is not installed at all. The diagram at the right indicates the
time to failure only starting from the installation time for each
unit (starting from $\bullet$, at the left). Unit 2 does not appear
on the right diagram as it has not been installed yet, while units 1,
4 and 6 are censored for $T$.} \label{fig2}
\end{figure}

\section{\texorpdfstring{Characterization of failure time.}{Characterization of failure time}}
So far our effort was to characterize the expected number of
unobserved installation times in different partitions of $[0,T_0]$.
Once this is known, we want to impute these installation times in an
iterative fashion (see Section \ref{chl}). For the time being, if we
assume the imputed samples represent the actual unobserved
installation times, it presents the case of random right censoring
for $T$.
This is explained in Figure~\ref{fig2}. The left-hand diagram in
Figure~\ref{fig2} represents the possible scenarios with both
installation and failure times. In the right-hand diagram of Figure~\ref{fig2}
we plot the time to failure for each unit, taking
installation time as the starting point. For the imputed
installation time (i.e., unobserved due to the fact that the unit is
still working) what we really get is $T_0 -X$ or the random
censoring time. Hence, the observed variable is $T^\ast=\min\{T,
T_0-X\}$. Note that $X$ and $T$ are assumed to be independent and so
are $T$ and $T_0 -X$. Let $\delta$ indicate whether $T^\ast$ is
censored ($\delta=0$) or it is a real failure ($\delta=1$). For the
current situation we have $C$ real failures and $[NF_X (T_0)-C]$
censored times, while $[N(1-F_X (T_0))]$ units do not contribute to
the estimation process as they provide no information related to
failure. The data from $n=[NF_X (T_0)]$ units consists of the pair
$(t_i^\ast,\delta_i)$. Since we are interested in inference about
the parameters of $F_T(\cdot)$, the likelihood function for the same is
given by
%
%
\begin{equation}\label{a6}
L_T=\prod_{i=1}^n [f_T (t_i^\ast)]^{\delta_i}
[S_T(t_i^\ast)]^{1-\delta_i}.
\end{equation}

\section{\texorpdfstring{Iterative algorithm.}{Iterative algorithm}}\label{chl}
All our earlier calculations are solely for the purpose of parameter
estimation in the distributions of $X$ and $T$. The key quantity of
the whole discussion is ${\alpha_{k+1}}$ (see Section \ref{sec3.1}), which
represents the number of unobserved installation times in
$[x_k,x_{k+1}]$. However, the estimation of ${\alpha_{k+1}}$
requires knowledge of the parameters in the distributions of $X$ and
$T$. We have assumed so far that the distributions of $X$ and $T$
are known; however, the parameters are actually unknown. Hence, an
iterative procedure is proposed.

\textit{Begin procedure}
\begin{enumerate}[Step 1.]
\item[Step 0.] Find initial parameter estimates of $F_X(\cdot)$ and
$F_T(\cdot)$ assuming that they are coming from a truncated distribution
($<T_0$) for which we have complete knowledge (e.g., Weibull,
Exponential, etc.).
\item[Step 1.] Using the current value of the distribution parameters, find
$\widehat\alpha_{k+1}$ for $k=0,\ldots, C$. Note that it is quite
possible to have $\widehat\alpha_{k+1}$ not as an integer, say,
$\widehat\alpha_{k+1}=\operatorname{int}(\widehat\alpha_{k+1})+\operatorname{frac}(\widehat\alpha
_{k+1})=U_{k+1}+V_{k+1}$.
\item[Step 2.] Draw $U_{k+1}$ samples from the interval $[x_k,
x_{k+1}]$ of the distribution $F_X(\cdot)$ using current values of the
distribution parameters.
\item[Step 3.] First, draw a sample from a $\operatorname{Bernoulli}(V_{k+1})$. If it is
equal to one, draw another sample as in step 2, otherwise skip to
the next step. Hence, the total number of imputed samples is either
$U_{k+1}$ or $U_{k+1}+1$.
\item[Step 4.] Re-estimate the parameters of $X$ using both imputed
and observed ($C$) samples via MLE under right censoring using
equation (\ref{a51}).
\item[Step 5.] Re-estimate the parameters of $T$ by using both
observed ($C$) and censored samples via equation (\ref{a6}). The
random censoring value for any imputed sample is $T_0-X_{\mathrm{imputed}}$.
\item[Step 6.] Return to step 1 until an acceptable
convergence tolerance level is reached on the parameter estimates.
\end{enumerate}

\textit{End procedure}

Note that the conventional approach stops at ``Step 0'' without any
further iteration, so we are simply using that as the initial guess.
Details for obtaining the MLE for some of the truncated
distributions (e.g., Exponential and Weibull) are described in the
\hyperref[app]{Appendix}. Though this algorithm assumes that the parametric
form of $X$ and $T$ are known, it does not depend upon any specific
distributional choice. Under the assumption that the specific
distributional choices of $F_X(\cdot)$ and $F_T(\cdot)$ are correct, the
speed of convergence depends upon the actual observed sample size
($C$) and end of study time ($T_0$). If $C$ is too small, it will
require many imputations (as $[NF_X (T_0)-C]$ is big). Similarly, if
$T_0$ is too small thus representing an early study termination, it
will force $C$ to be quite small. Both of these cases represent very
little available information. This generally results in large
sampling variance with high fluctuations in the iterations resulting
in nonconvergence.

\section{\texorpdfstring{Connection with the exact likelihood.}{Connection with the exact likelihood}}\label{sec6}
Note that our main goal is to estimate parameters in the
distribution of $X$ and $T$ and typically a likelihood is a function
of those parameters. As noted earlier in Section \ref{sec3}, though $X$ and
$T$ are assumed to be independent, the nature of \textit{ambiguous
censoring} make their joint distribution dependent, where the
functional component related to respective parameters are
nonseparable. As a consequence, maximum likelihood estimation
requires joint maximization for all parameters over the exact
likelihood function given in equation (\ref{e2}), which is
computationally prohibitive. Thus, a major point in this article is
the separation of the $X$ and $T$ distributions via equations
(\ref{a51}) and~(\ref{a6}). A pertinent question is the theoretical
justification of the above in light of the exact likelihood. Note
that $P\{X+T>T_0\}=S_X (T_0) + \int_0^{T_0} S_T (T_0-x)\,dF_X (x)$. In
case there is an oracle which supplies us information about the
$N-C$ unobserved units, that is, whether $\{X>T_0\}$ or $\{X\leq
T_0\}\cap\{T>T_0 - X\}$, the above expression simplifies
considerably. Suppose that out of those $N-C$ units we know that
$|\Gamma|$ ($\simeq[NF_X (T_0)-C]$) units are installed (with
reported installation times) but have not yet failed by $T_0$; then
for those units, $P\{X+T>T_0\}=f_X(x)S_T(T_0-x)$. For the remaining
$N-|\Omega|-|\Gamma|$ ($\simeq[N(1-F_X (T_0))]$) no information is
available, as they are not installed. Hence, we get type-1 right
censoring on $X$ at $T_0$, implying $P\{X+T>T_0\}=S_X (T_0)$. The
likelihood contribution from the imputed and unobserved units is
$\{\prod_{j\in\Gamma}f_X(x_j)
S_T(T_0-x_j)\}S_X(T_0)^{N-|\Omega|-|\Gamma|}$. Under the above
setup, the complete likelihood for all observed and imputed samples
becomes
%
%
\begin{eqnarray}\label{e3}
&&L(F_X, F_T) \nonumber\\
&&\quad\propto \biggl\{ \prod_{i\in\Omega} f_X(x_i)
f_T(t_i)\biggr\}
\biggl\{ \prod_{j\in\Gamma}f_X(x_j)S_T(T_0-x_j)\biggr\}
S_X(T_0)^{N-|\Omega|-|\Gamma|}\\
&&\quad\propto \biggl\{S_X(T_0)^{N-|\Omega|-|\Gamma|} \prod_{i\in
\Omega\cup\Gamma} f_X(x_i) \biggr\} \biggl\{\prod_{i\in
\Omega}f_T(t_i) \prod_{j\in\Gamma}S_T(T_0-x_j) \biggr\}.\nonumber
\end{eqnarray}
This is what corresponds to equations (\ref{a51}) and (\ref{a6}).
%

\section{\texorpdfstring{Simulation studies.}{Simulation studies}}\label{sec7}
Next we present some simulation studies with different choices of
reliability distributions to demonstrate the efficacy of the
proposed approach. In particular, we consider exponential and
Weibull distributions for both~$X$ and $T$ with different values of
$T_0$. To explain the convergence criteria let us assume $\mu$ is a
parameter (in either $X$ or $T$) that needs to be estimated. We stop
the iteration when $|\frac{\mu_{i+p} - \mu_{i}}{\mu_{i+p}
}|<\varepsilon$, where $i$ denotes the iteration number, $p$ is a
prespecified positive integer constant and $\varepsilon$ is a
prespecified small value chosen by the end user. For
multi-parameter cases this needs to be satisfied for every
parameter. Alternatively, in the spirit of the Monte-Carlo-based
approach, we may run a fixed but large number of iterations and
discard the first few iterations as nonstabilized (or ``burn-in'')
values and keep all the remaining to report the estimated empirical
mean and standard deviation. We took the second approach as we found
that convergence is very fast even for $\varepsilon=0.0005$, except for
the situation when $\frac{C}{N}<20\%$. In every situation we also
report the exact stopping time if we choose to use the first
stopping criterion (i.e., stop if $|\frac{\mu_{i+p} -
\mu_{i}}{\mu_{i+p} }|<\varepsilon$). We also report the exact runtime
in every simulation using R code on a Windows-XP-based machine until
convergence. We hope this should give the reader a comprehensive
idea about the run time efficacy of our approach. The computer code
used for the simulation is available as a supplementary material
[\citet{ghosh9}].

%
\begin{sidewaystable}
\caption{The simulation result $N=200$. $C$ denotes total observed
samples, while $|D|$ denotes true unobserved installations before
$T_0$ } \label{simul1}
\begin{tabular*}{\textwidth}{@{\extracolsep{\fill}}lcccccccc@{}}
\hline
\textbf{Different} &  &  &  & \textbf{Initial} & \textbf{Simulation} & \textbf{Average No.} &
\textbf{Convergence} & \textbf{Time in}\\
\textbf{distribution} & $\bolds{T_0}$& $\bolds{C}$& $\bolds{|D|}$& \textbf{estimates} & \textbf{results} & \textbf{imputations} & $\bolds{p=5,}$
$\bolds{\varepsilon=0.0005}$ & \textbf{second}
\\
\hline
$X\sim \operatorname{Exp}(\lambda=0.2)$ & 6 & \phantom{0}75 & \phantom{0}67 & $\lambda=0.43$ &
$\widehat{\lambda}=0.19$, $\widehat\sigma_\lambda=0.021$ & 60
& \phantom{0}57 & 121 \\
$T\sim \operatorname{Exp}(\delta=0.2)$& & & & $\delta=0.51$ &
$\widehat{\delta}=0.23$, $\widehat\sigma_\delta=0.027$ & & & \\
$X\sim \operatorname{Exp}(\lambda=0.2)$ & 5 & \phantom{0}47 & \phantom{0}75 & $\lambda=0.5$ &
$\widehat{\lambda}=0.18$, $\widehat\sigma_\lambda=0.026$ & 72
& \phantom{0}46 & \phantom{0}97 \\
$T\sim \operatorname{Exp}(\delta=0.2)$& & & & $\delta=0.48$ &
$\widehat{\delta}=0.19$, $\widehat\sigma_\delta=0.029$ & & & \\
$X\sim \operatorname{Exp}(\lambda=0.5)$ & 6 & 108 & \phantom{0}84 & $\lambda=0.69$ &
$\widehat{\lambda}=0.48$, $\widehat\sigma_\lambda=0.018$ & 78
& \phantom{0}32 & 133 \\
$T\sim \operatorname{Exp}(\delta=0.2)$& & & & $\delta=0.4$ &
$\widehat{\delta}=0.22$, $\widehat\sigma_\delta=0.01$ & & & \\
$X\sim \operatorname{Exp}(\lambda=0.5)$ & 4 & \phantom{0}66 & 102 & $\lambda=0.81$ &
$\widehat{\lambda}=0.43$, $\widehat\sigma_\lambda=0.025$ & 98
& \phantom{0}65 & 116 \\
$T\sim \operatorname{Exp}(\delta=0.2)$& & & & $\delta=0.56$ &
$\widehat{\delta}=0.23$, $\widehat\sigma_\delta=0.013$ & & & \\
$X\sim \operatorname{Exp}(\lambda=0.4)$ & 6 & 170 & \phantom{0}13 & $\lambda_X=0.53$ &
$\widehat{\lambda}=0.44$, $\widehat{\sigma}=0.013$ & 18
& 101 & 108 \\
$T\sim \operatorname{Exp}(\delta=0.7)$& & & & $\delta=0.78$ &
$\widehat{\delta}=0.7$, $\widehat\sigma_\delta=0.03$ & & & \\
$X\sim \operatorname{Exp}(\lambda=0.4)$ & 4 & 124 & \phantom{0}43 & $\lambda=0.67$ &
$\widehat{\lambda}=0.41$, $\widehat{\sigma}=0.03$ & 36
& 212 & 155 \\
$T\sim \operatorname{Exp}(\delta=0.7)$& & & & $\delta=1.07$ &
$\widehat{\delta}=0.75$, $\widehat\sigma_\delta=0.06$ & & & \\
$X\sim \operatorname{Exp}(\lambda=0.7)$ & 6 & 111 & \phantom{0}84 & $\lambda=1.05$ &
$\widehat{\lambda}=0.66$, $\widehat\sigma_\lambda=0.03$ & 88
& \phantom{0}66 & 123 \\
$T\sim \operatorname{Weibull}(\beta=2,\theta=5)$& & & & $\beta=1.21$&
$\widehat{\beta}=2.03$, $\widehat\sigma_\beta=0.04$ & & & \\
$\beta=\mathrm{Shape}$, $\theta=\mathrm{Scale}$ & & & & $\theta=1.71E+03$ &
$\widehat{\theta}=5.04$, $\widehat\sigma_\theta=0.14$& & &
\\
$X\sim \operatorname{Weibull}(\beta=1.5,\theta=4)$& $6$ & 107 & \phantom{0}47 &
$\beta=2.51$&
$\widehat{\beta}=1.51$, $\widehat\sigma_\beta=0.04$ & 35 & \phantom{0}45 &115 \\
& & & & $\theta=5.15$ & $\widehat{\theta}=3.38$,
$\widehat\sigma_\theta=0.14$& & & \\
$T\sim \operatorname{Exp}(\lambda=0.5)$ & & & & $\lambda=0.74$ &
$\widehat{\lambda}=0.44$, $\widehat\sigma_\lambda=0.033$& & &
\\
\hline
\end{tabular*}
\end{sidewaystable}
%
\begin{figure}
\centering
\begin{tabular}{@{}c@{\quad}c@{}}

\includegraphics{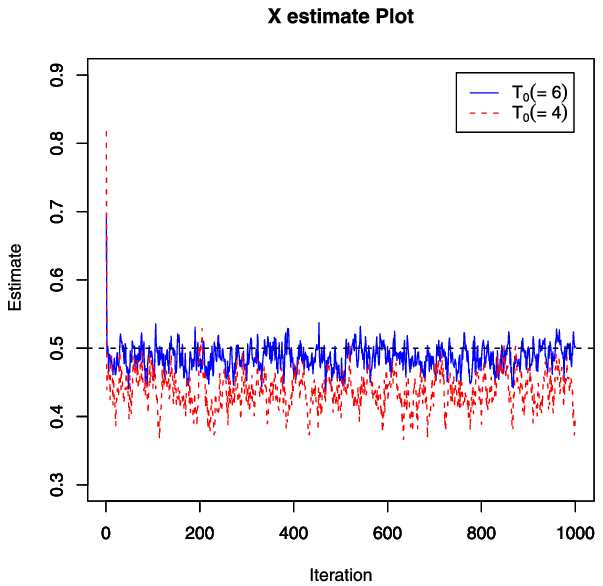}
 & \includegraphics{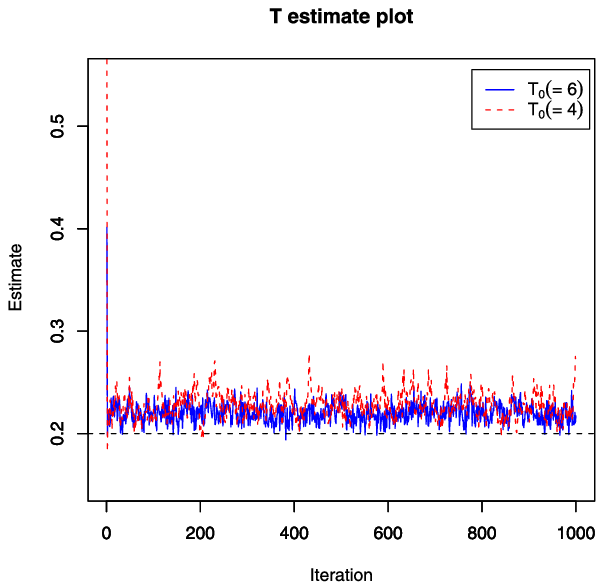}\\
\footnotesize{(a)} & \footnotesize{(b)}\\[3pt]

\includegraphics{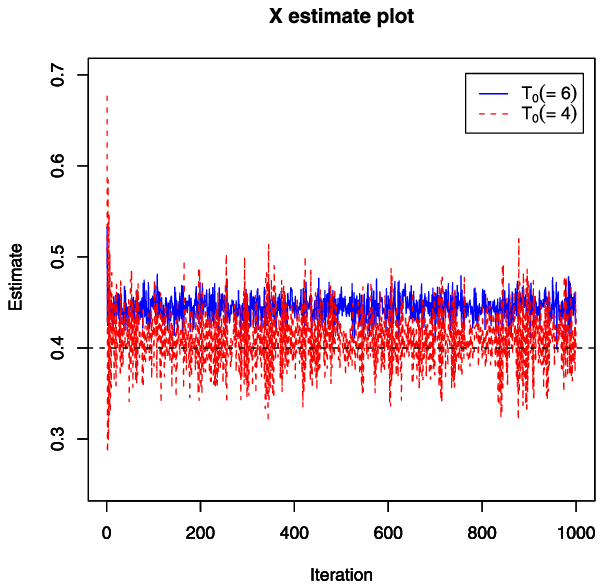}
 & \includegraphics{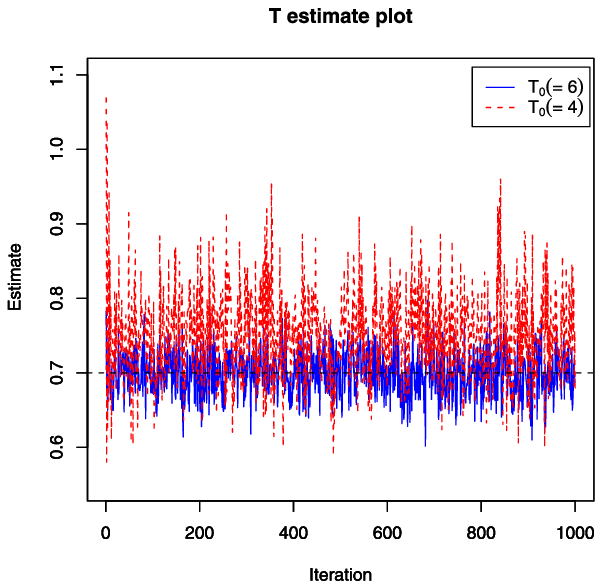}\\
\footnotesize{(c)} & \footnotesize{(d)}
\end{tabular}
\caption{Plot of the maximum likelihood estimate over different
iterations for nonidentical exponential cases. Plots at the top are
for $\lambda=0.5$, $\delta=0.2$ and at the bottom are for
$\lambda=0.4$, $\delta=0.7$. The ``$\cdots$'' (dashed line) indicates
the true value of the parameter in each case.}\label{fig3}
\end{figure}

%
\begin{figure}
\centering
\begin{tabular}{@{}c@{\quad}c@{}}

\includegraphics{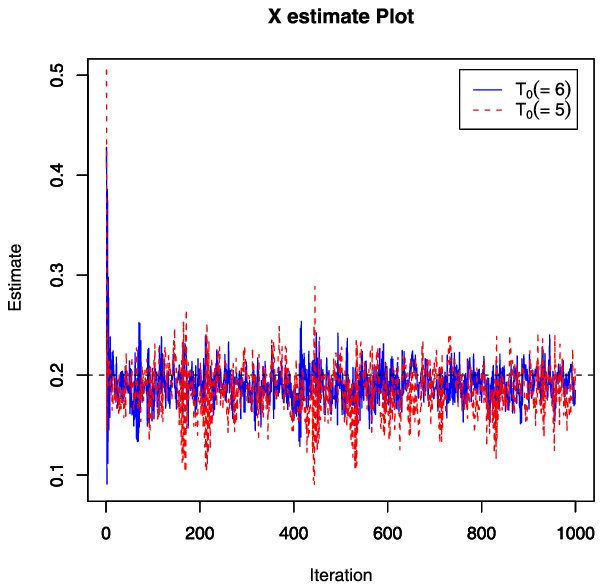}
 & \includegraphics{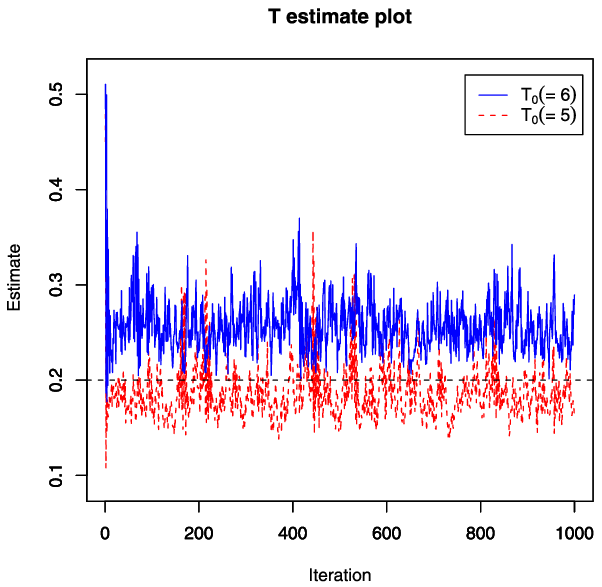}\\
\footnotesize{(a)} & \footnotesize{(b)}\\
\end{tabular}
\caption{Plot of the maximum likelihood estimate over different
iterations for the i.i.d. exponential case ($\lambda=\delta=0.2$).
The ``$\cdots$'' (dashed line) indicates true value of the parameter
in each case.}\label{fig31}
\end{figure}

%
\begin{figure}
\centering
\begin{tabular}{@{}cc@{}}

\includegraphics{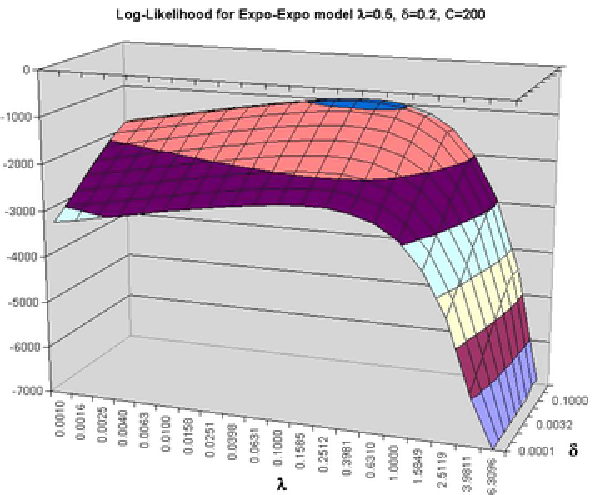}
 & \includegraphics{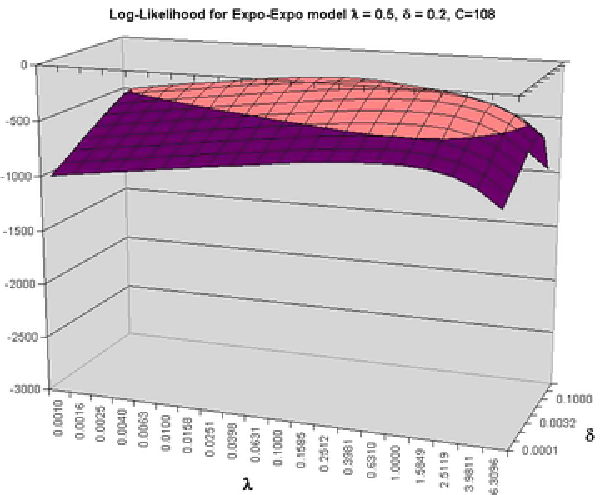}\\
\footnotesize{(a)} & \footnotesize{(b)}\\

\includegraphics{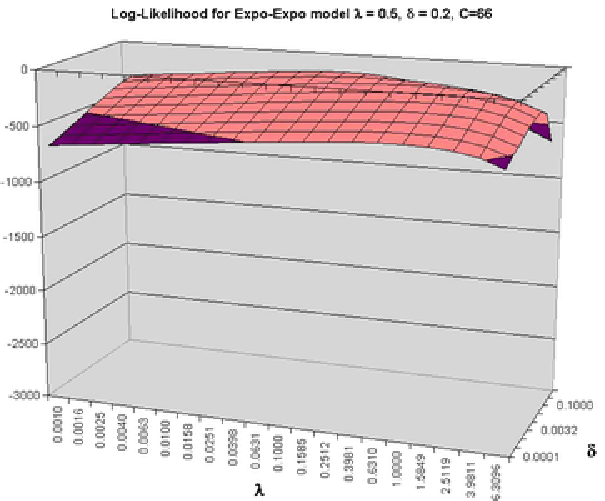}
 & \includegraphics{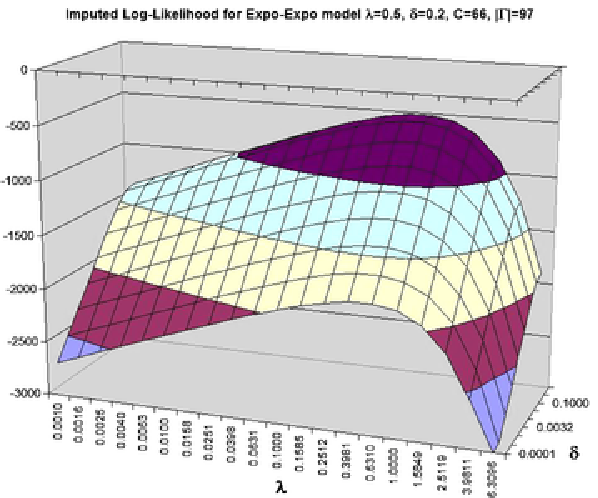}\\
\footnotesize{(c)} & \footnotesize{(d)}
\end{tabular}
\caption{Log-likelihood surface plots for the
Exponential--Exponential model with $\lambda=0.5$ and $\delta=0.2$.
(\textup{a}) is obtained when we use all the observations ($C=200$).
(\textup{b}) is obtained when $T_0=6$ and $C=108$. (\textup{c}) is
obtained when $T_0=4$ and $C=66$. Likelihood becomes flatter as
$C\downarrow$, thus making MLE search a difficult task. (\textup{d})
is obtained for a specific iteration when imputation is used ($97$
imputed samples) for $T_0=4$ and $C=66$.} \label{fig8}
\end{figure}

\begin{figure}
\centering
\begin{tabular}{@{}c@{\hspace*{1pt}}c@{}}

\includegraphics{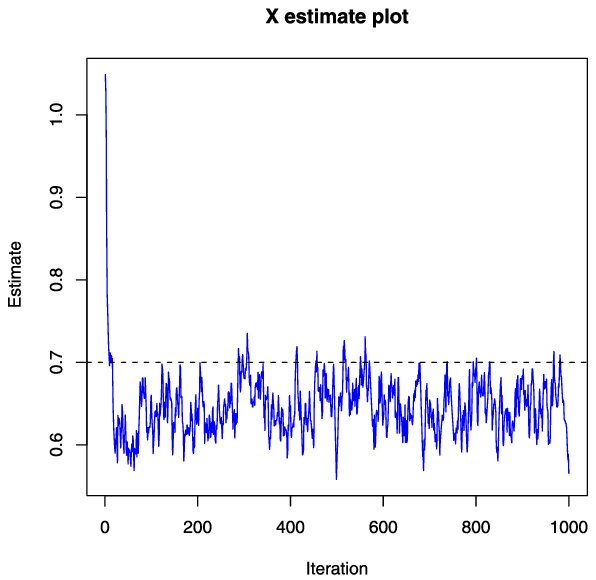}
 & \includegraphics{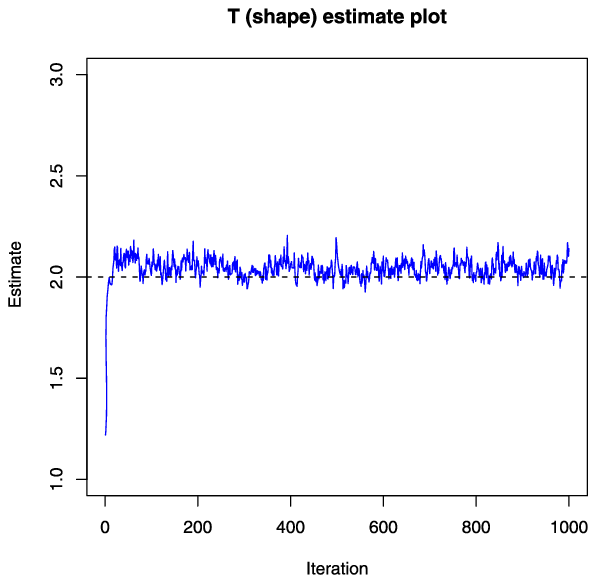}\\
\footnotesize{(a)} & \footnotesize{(b)}
\end{tabular}\vspace*{3pt}
\begin{tabular}{@{}c@{}}

\includegraphics{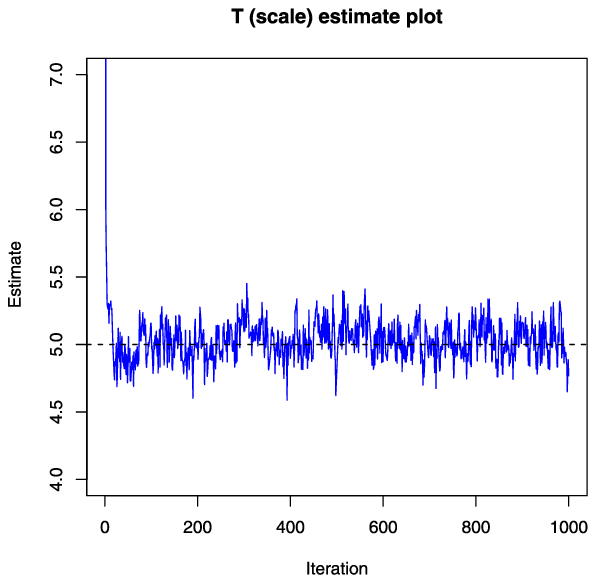}
\\
\footnotesize{(c)}
\end{tabular}
\caption{Plot of the maximum likelihood estimate for the
Exponential--Weibull model over different iterations. The ``$\cdots$''
(dashed line) indicates true value of the parameter in each
case.}\label{fig4}
\end{figure}

\begin{figure}
\centering
\begin{tabular}{@{}c@{\hspace*{1pt}}c@{}}

\includegraphics{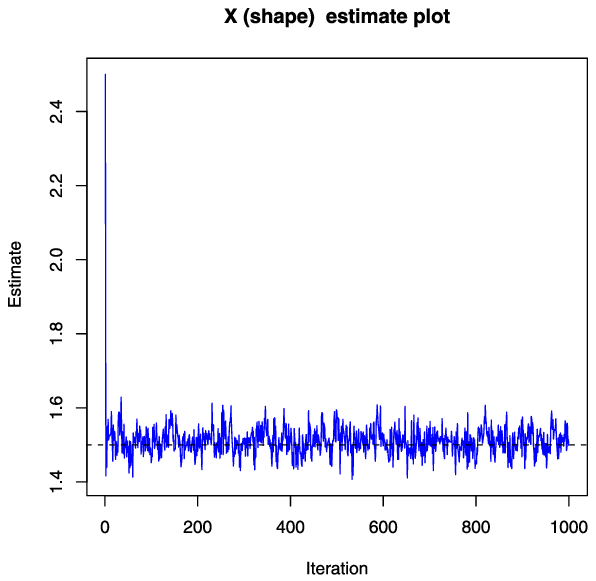}
 & \includegraphics{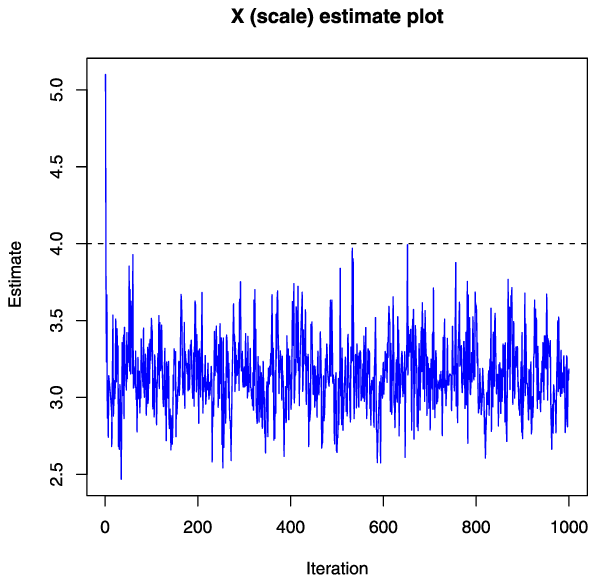}\\
\footnotesize{(a)} & \footnotesize{(b)}
\end{tabular}\vspace*{3pt}
\begin{tabular}{@{}c@{}}

\includegraphics{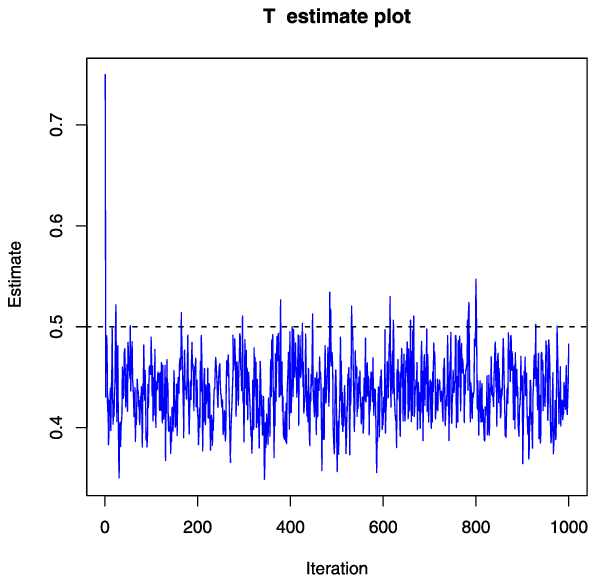}
\\
\footnotesize{(c)}
\end{tabular}
\caption{Plot of the maximum likelihood estimate for the
Weibull--Exponential model over different iterations. The ``$\cdots$''
(dashed line) indicates true value of the parameter in each
case.}\label{fig5}
\end{figure}

Table \ref{simul1} represents the simulation results for different
choices of distributions for $X$ and $T$. We choose $N=200$ for all
experiments. We run the iteration $1000$ times for each model, of
which we discard the first $100$ as burn-in values. The reported
parameter estimates and standard deviations are based on the
remaining $900$ iterations. We also report the convergence iteration
number, which, for the multi-parameter case, represents the maximum
of all iterations taken by individual parameters to satisfy
$|\frac{\mu_{i+p} - \mu_{i}}{\mu_{i+p} }|<\varepsilon$. As we can see
from Table \ref{simul1}, convergence is achieved quickly. For
parameter estimation we used the maximum likelihood approach which
is described briefly in the \hyperref[app]{Appendix} section. Again for other
nontrivial distributions with complicated MLE, the method of
moments or rank regression based approaches [\citet{john}] could be
used. In each model, following standard practice, we obtain the
initial parameter estimates for the distribution of $X$ and~$T$
using the right truncated distribution. These initial estimates are
way off in all cases, which explains why standard practice is
unsatisfactory in this nontrivial situation. We summarize our
simulation result in Table \ref{simul1}. The first two rows in Table
\ref{simul1} are of special interest since we assumed
$X,T\stackrel{\mathrm{i.i.d.}}{\sim}\operatorname{Exp}(\lambda=\delta)$. As shown in
\hyperref[app]{Appendix B.5}, the exact likelihood given in equation (\ref{e2}) can be solved
numerically in this case. For $T_0=6$ the exact likelihood based MLE
yields $\widehat{\lambda}=0.22$ with asymptotic standard deviation
$\widehat\sigma_\lambda=0.026$. For $T_0=5$, we get
$\widehat{\lambda}=0.18$ with asymptotic standard deviation
$\widehat\sigma_\lambda=0.028$. In both of these cases our
simulation result is very close to the true value
($\lambda=\delta=0.2$) even though we did not use the information
that $\lambda=\delta$ in our proposed algorithm. In Figure
\ref{fig31} we present pictorially the result for these two cases.
This supports the viability of our algorithm. Next we explore
non-i.i.d. cases. Figure \ref{fig3} presents the case for $X\sim
\operatorname{Exp}(\lambda)$ and $T\sim \operatorname{Exp}(\delta)$ with two different observation
times ($T_0=4, 6$). In the first case, we choose the true model
parameters in such a way that about $50\%$ of the cases are observed
(i.e., $C>100$). Figure \ref{fig3}(a) and (b) present the case when
$\lambda=0.5$ and $\delta=0.2$. We observe $108$ and $66$ units for
$T_0=6$ and $4$, respectively. As expected, the case with more units
produces better estimates. Nevertheless, we point out that for
$T_0=4$, even though we observe only about $33\%$ of the units, the
final parameter estimates are still noticeably close to the true
parameter values. Similar observations could be made for the other
choice of parameter values in Figure \ref{fig3}(c) and (d). To
elucidate the problem when using the exact maximum likelihood based
approach, we have also plotted the log-likelihood surface (obtained
via equation (\ref{e2}) and numerical integration) in Figure
\ref{fig8} for the case $\lambda=0.5$ and $\delta=0.2$. Figure
\ref{fig8}(a) represents the case when we have complete
observations for all units ($C=N$). However, as~$T_0$ shrinks, $C$
goes down, and, as a result, the likelihood surface becomes very flat.
Hence, searching for the MLE becomes computationally challenging and
often leads to large variance. We have noted this problem earlier in
Section~\ref{sec2}. Figure~\ref{fig8}(d) presents the log-likelihood surface
obtained via equation (\ref{e3}) when imputation is in use. This
representative plot is obtained for a specific iteration when $97$
units are imputed while running the algorithm described in Section
\ref{chl}. The flatness of the resulting log-likelihood surfaces in Figure
\ref{fig8}(c) and (d) is an indicator of computational difficulties
in finding the MLE for each case. Next, in Figure~\ref{fig4} we
describe the iteration result when $X\sim \operatorname{Exp}(\lambda)$ and $T\sim
\operatorname{Weibull}(\beta,\theta)$. In Figure~\ref{fig5} we describe the
iteration result when $X\sim \operatorname{Weibull}(\beta,\theta)$ and $T\sim
\operatorname{Exp}(\lambda)$. In all cases the final estimates are quite close to
the true model parameters. Though not reported here, we obtain
similar results with the gamma distribution. For details of the
sampling from a truncated gamma distribution, please refer to
\citet{damien}. We have confined our simulation exploration only to
commonly used reliability distributions; however, we are hopeful
that the algorithm presented here will also work for other
distributions with nonnegative support.

%

%
\begin{figure}
\centering
\begin{tabular}{@{}c@{\hspace*{4pt}}c@{}}

\includegraphics{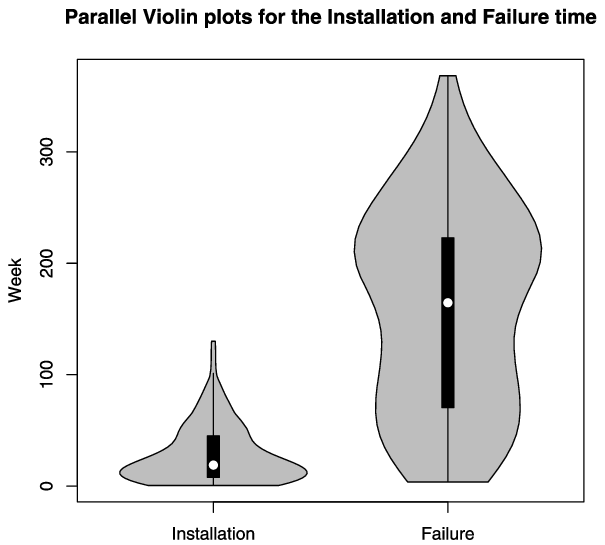}
 & \includegraphics{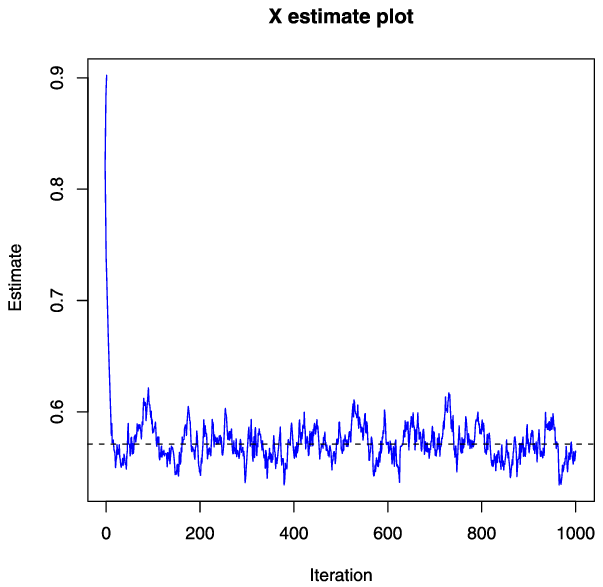}\\
\footnotesize{(a)} & \footnotesize{(b)}\\[3pt]

\includegraphics{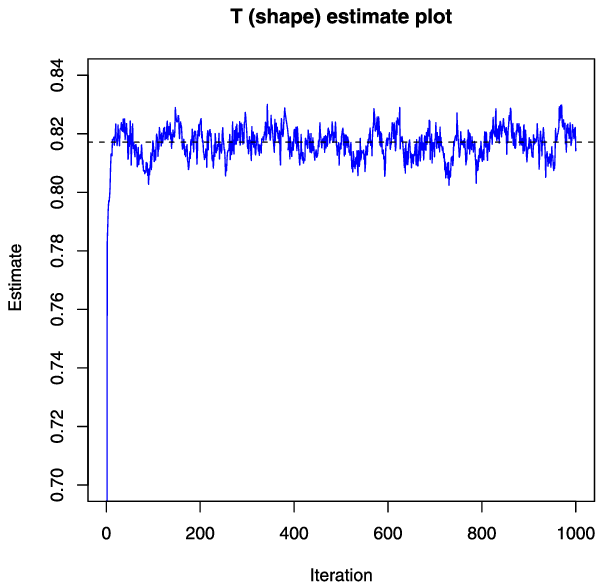}
 & \includegraphics{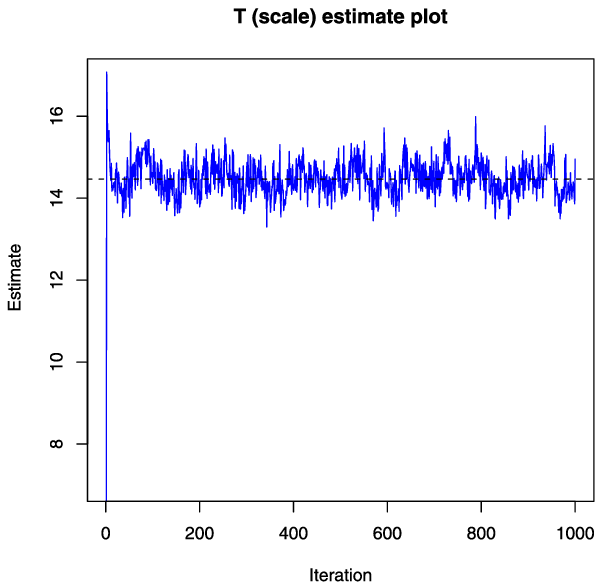}\\
\footnotesize{(c)} & \footnotesize{(d)}
\end{tabular}
\caption{Plot (\textup{a}) represents violin plot for the observed 133 units.
Other plots represent the maximum likelihood estimates for the
Furnace data over different iterations. The ``$\cdots$'' (dashed
line) indicates finally estimated mean of the parameter in each
case.}\label{fig6}
\end{figure}

\section{\texorpdfstring{Motivating application.}{Motivating application}}\label{motiv}
\begin{figure}
\centering
\begin{tabular}{@{}cc@{}}

\includegraphics{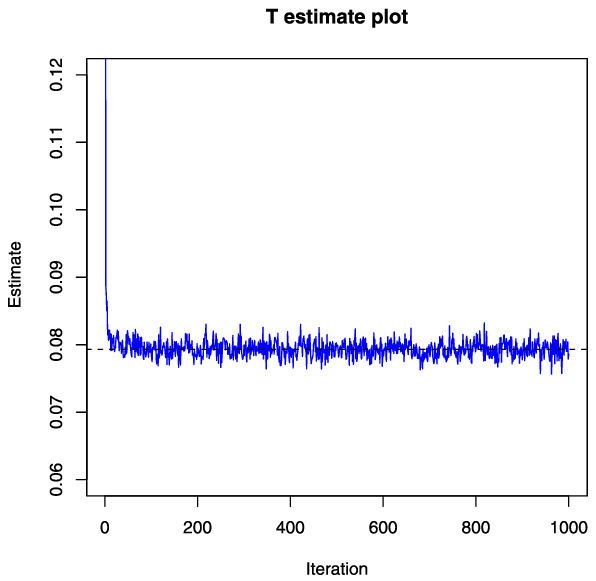}
 & \includegraphics{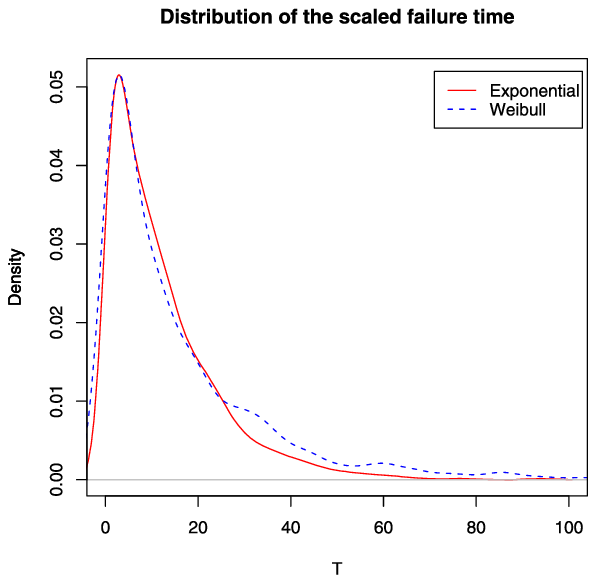}\\
\footnotesize{(a)} & \footnotesize{(b)}
\end{tabular}\vspace*{3pt}
\begin{tabular}{@{}c@{}}

\includegraphics{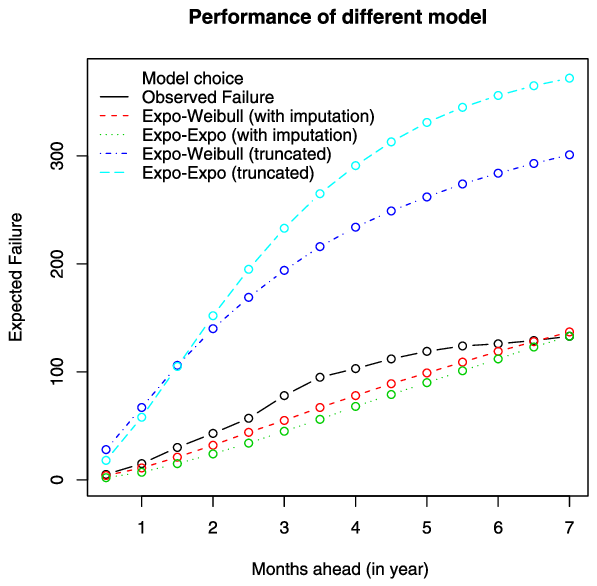}
\\
\footnotesize{(c)}
\end{tabular}
\caption{On the left, (\textup{a}), maximum likelihood estimate for the
furnace data when $T\sim \mathit{Exponential}$ distribution. In the middle,
(\textup{b}), density plot for two different model choices for $T$. Both look
similar. On the right, (\textup{c}), it represents performance of different
models compared with the observed failure. Truncated distribution
with no imputation performs very poorly with huge overestimation.
Performance of imputed models are far better and the Exponential--Weibull
model choice does the best job.}\label{fig7}
\end{figure}

The data set that we will analyze using the current procedure came
from an industrial house producing residential furnace components
during one week in May $2001$. We consider a batch with $N=400$
units. The data consist of $C=133$ pairs of points as observed units
(i.e., $\{x_i, t_i\}_{i=1}^{133}$), which have failed within the
observation time of seven years from the date of manufacturing.
Figure \ref{fig6}(a) shows a violin plot for installation and
failure times. The violin plot is a combination of a box plot and a
kernel density plot. There is no specific information available
about the remaining units. We are assuming that there exists no unit
which has failed but was not reported. In practice, this could have
happened for many other reasons. In the present context the
reliability engineers believe that it is appropriate to model
installation time ($X$) using an exponential distribution, while
failure time ($T$) is modeled according to a Weibull distribution [\citet{jager}; \citet{xiao}].
It should be noted that seasonality plays an important role
in selling, installation and duty cycles (how rigorously the unit is
being used) of the product. However, since in the present case we
consider only a single batch, we assume that these effects will be
similar for every unit in the batch. When comparing the units
produced under different batches (and possibly produced at different
times of the year), additional care is required as the independence
assumption between $X$ and $T$ becomes questionable. This is due to
the fact that some installation times are associated with severe
duty cycles and more reliability problems.
Before running the algorithm we divide the installation times as
well as failure times by their corresponding standard deviation
estimated from $133$ samples. This rescaling is done for numerical
stabilization only, which results in faster convergence of the
algorithm. Rescaled random variables have straightforward
relationships with the original variables, without any drastic
change to the distributional form. We run the algorithm for $1000$
iterations, however, convergence (with $p=5$, $\varepsilon=0.0005$) was
achieved much earlier. We discard the first $100$ iterations as
burn-in and report the estimates on the basis of the remaining $900$
iterations in Table~\ref{simul2}. For model comparison purposes we
have also investigated separately the case where $T$ is assumed to
follow the exponential distribution, without altering the
distribution of $X$. In each case we obtain the initial parameter
estimates using the right truncated distributions. Figure \ref{fig6}
represents the case for the Exponential--Weibull model combination.
Though the Exponential--Exponential model parameter is different from
the previous choice (see Table \ref{simul2}), the density plot of
the two distributions of $T$ are quite similar as depicted in Figure
\ref{fig7}(b). We have also compared the predictive performance of
different models in Figure \ref{fig7}(c), including the usual
practice of truncated distributions without any imputation. We
estimated the expected number of failures to be observed for
different observation times over an interval of six months. This
expected failure number is then compared with the observed failure
number for the current data set. This required repeated
re-estimation of model parameters at different time points. As can
be seen, the truncated models have a huge overestimation problem
throughout the study period. This again justifies our earlier
criticism of current practice. Imputed models produce stable
estimates and do much better even at the very early stage of product
lifetime with only limited data. The Exponential--Weibull model
choice does a little better than the Exponential--Exponential model.
However, they are very much comparable as expected from Figure
\ref{fig7}(b). It is desirable to estimate the expected failure
number accurately for two main reasons. First, by accurately
estimating warranty claims, an estimate of required financial
reserves can be performed. This has immense implications in terms of
future financial resource management. Second, it is desired to
continuously improve the quality of consumer products, especially at
the very high quality levels enjoyed by many consumer products
today. All these aspects necessarily depend upon the accurate and
efficient estimation of the reliability parameters (in $X$ and $T$).
The method described in this paper provides a first step in this
direction.

%
\begin{table}
\tabcolsep=0pt
\caption{Estimates for $N=400$ units in a single batch. $C=133$
units have complete observations} \label{simul2}
\begin{tabular*}{\textwidth}{@{\extracolsep{\fill}}lccccc@{}}
\hline
 & \textbf{Initial} & \textbf{Simulation} & \textbf{Average No.} & \textbf{Convergence} & \textbf{Time in}\\
\textbf{Distribution}& \textbf{estimate} & \textbf{result} & \textbf{imputations} & \textbf{iteration} & \textbf{second}\\
\hline
$X\sim \operatorname{Exp}(\lambda)$ & $\lambda=0.9$ & $\widehat{\lambda}=0.57$,
$\widehat\sigma_\lambda=0.014$ & $260$ & $167$ & 381\\
$T\sim \operatorname{Weibull}(\beta, \theta)$ & $\beta=0.6$,  &
$\widehat{\beta}=0.81$, $\widehat\sigma_\beta=0.004$ & & &
\\
& $\theta=3.18$& $\widehat{\theta}=14.47$, $\widehat\sigma_\theta=0.4$ & & & \\
$T\sim \operatorname{Exp}(\delta)$ & $\delta=0.51$&
$\widehat{\delta}=0.079$,
$\widehat\sigma_\delta=0.001$ & $263$ & \phantom{0}45 & 421\\
\hline
\end{tabular*}
\end{table}

%

%
\section{\texorpdfstring{Concluding remarks.}{Concluding remarks}}
Unlike electronic commodities, item specific tracking is not a
feasible solution for many large scale industrial operations. Hence,
the availability of both ``complete'' and ``partial'' information is
quite common. In addition, except for very rare occasions, there are
hardly any situations where all units in a batch start working at
the same time. Unavailability of the installation time in a timely
fashion is a major challenge to reliability engineers. Because of
confidentiality issues we can not reveal any company specific
information. However, we would like to mention that the above problem
exists in different industrial sectors, and there is no clear
solution thus far. In this paper we have proposed a computational
approach to solve the problem with the optimal usage of partial and
complete information. From a reliability engineer's perspective,
this current approach is simple, fast and also has straightforward
interpretability.

The primary focus of any reliability analysis is the failure time.
However, the waiting time for the installation is also very important
in the sense that it provides valuable market specific information
from the sales perspective, including seasonality and periodic sales
patterns. In our approach we have targeted simultaneous estimation
for both installation and failure time parameters in a combined
fashion. To the best of our knowledge, this is the first attempt to
do so. Finally, we would like to point out some of the assumptions
that we have made in this paper, a violation of which will require
more research. First, we have assumed that installation time and
failure time are independent. This may be questionable in some
situations as discussed in Section \ref{motiv}. Second, there is no aging
effect for the units installed at different time points. Finally, we
made the assumption that the distributional form of both
installation and failure times is known. While for most of the
legacy industrial products, in-house experts have a good idea about
this from historical knowledge, it is of theoretical interest to see
the effect of convergence and the quality of parameter estimates
under incorrect parametric model specification. One way to avoid
this is to choose a larger class of models. From the reliability
perspective there is considerable effort to generalize Weibull and
other popular reliability distributions [see \citet{bali} and
\citet{shao}]. However, the resultant estimation procedure will be
more involved. Another possibility is a nonparametric extension;
however, the resulting procedure will be much more complex. In an
ongoing work we are also exploring the exact probabilistic and
inferential procedure based on equation (\ref{e2}).

\begin{appendix}

\section{\texorpdfstring{Proof of Theorem \protect\lowercase{\ref{t2}}.}{Proof of Theorem 3.2}}
We can use the inequality (\ref{a3}) to argue that the following
holds:
\begin{eqnarray*}
S_T (T_0 -x_k)&\leq& P[T>T_0 - X|x_k<X<x_{k+1}] \leq S_T (T_0
-x_{k+1}),
\\
S_T (T_0 -x_{k-1})&\leq& P[T>T_0 - X|x_{k-1}<X<x_k] \leq S_T (T_0
-x_{k}).
\end{eqnarray*}
Combining both of these yields the proof.

\section{\texorpdfstring{Maximum likelihood estimation.}{Maximum likelihood estimation}}\label{app}
We concentrate here on Exponential and Weibull distribution as used
in the simulation, though other distributions with positive support,
such as gamma and log-normal, can also be considered. Most of the
results are published elsewhere and referenced as required.

\subsection{\texorpdfstring{Truncated exponential.}{Truncated exponential}}
Let $X\sim \operatorname{Exp}(\lambda)$ with $0\leq X \leq T_0$. The p.d.f. is given
by
\[
f(x|\lambda, T_0)=\frac{ \lambda\exp(-x\lambda)}{1-\exp{(-T_0
\lambda)}}.
\]
If we have $n$ observations, then differentiating the log-likelihood
equation with respect to $\lambda$ and equating it to zero yields
\[
\frac{1}{\lambda}-\frac{T_0 \exp{(-T_0 \lambda)}}{1-\exp{(-T_0
\lambda)}}-\overline{x}=0.
\]
The above equation needs to be solved numerically to get the MLE of
$\lambda$.
\subsection{\texorpdfstring{Randomly right censored exponential.}{Randomly right censored exponential}}
Let $T\sim \operatorname{Exp}(\lambda)$ and we observe $T^\ast=\min\{T, C_r\}$,
where in the current context $C_r=T_0-X$ and $X$ is another random
variable denoting installation time. Let us denote our samples as
$\{t_i^\ast, \delta_i\}_{i=1}^n$, where $\delta_i=1$ means the
sample is an actual observation and $0$ means it is censored. If we
have $\sum_{i=1}^n \delta_i=C$ true observations, then the
log-likelihood is given by
\[
L(\lambda)=c\log\lambda-\lambda\sum_{i=1}^C t_i -
\lambda\sum_{j=1}^{n-C} (T_0 - x_j),
\]
which upon equating to $0$ yields
$\widehat{\lambda}=\frac{C}{\sum_{i=1}^C t_i +\sum_{j=1}^{n-C} (T_0
- x_j)}$.
\subsection{\texorpdfstring{Truncated Weibull.}{Truncated Weibull}}
The MLE calculation for the truncated Weibull distribution is
somewhat involved and may not always exist. Some explicit
mathematical formulations with the required regularity conditions are
described in \citet{mittal}. We briefly mention only the final
result here that has been used in this paper. Suppose $X\sim
\operatorname{Weibull}(\beta,\theta)$, but with $0\leq X \leq T_0$. Let us denote
by $Y=\frac{X}{T_0}$. Unfortunately, the MLE for $\beta$ is not
available in closed form and needs to be solved numerically using
the equation
\[
\frac{\sum_{i=1}^n y_i^\beta}{n}- \frac{\sum_{i=1}^n y_i^\beta\log
y_i}{{n}/{\beta}+\sum_{i=1}^n \log y_i}+\biggl[
\exp\biggl\{\frac{{n}/{\beta}+\sum_{i=1}^n \log y_i}{\sum_{i=1}^n
y_i^\beta\log y_i}
\biggr\}-1\biggr]^{-1}=0.
\]
Once we know $\widehat{\beta}$, the MLE of $\theta$ is
\[
\widehat{\theta}=T_0\biggl(\frac{\sum_{i=1}^n y_i^{\widehat{\beta}}
\log y_i}{{n}/{\widehat{\beta}}+\sum_{i=1}^n \log
y_i}\biggr)^{{1}/{\widehat{\beta}}}.
\]
%
\subsection{\texorpdfstring{Randomly right censored Weibull.}{Randomly right censored Weibull}}
Suppose $T\sim \operatorname{Weibull}(\beta,\theta)$. Similar to the randomly right
censored exponential case $T^\ast=\min\{T, C_r\}$, where in the
current context $C_r=T_0-X$. We denote our data set as $\{t_i^\ast,
\delta_i\}_{i=1}^n$ and $\sum_{i=1}^n \delta_i=C$. The MLE is given
explicitly in \citet{shao} and \citet{lemon}, which again needs to be solved
numerically for $\beta$ using the equation
\[
\frac{1}{{\beta}}+\frac{\sum_{i=1}^n \delta_i \operatorname{log}
t_i^\ast}{C}-\frac{\sum_{i=1}^n (t_i^\ast)^{{\beta}} \log
t_i^\ast}{\sum_{i=1}^n (t_i^\ast)^{{\beta}}}=0.
\]
Once we know $\widehat{\beta}$, the MLE of $\theta$ is
\[
\widehat{\theta}=\biggl( \frac{\sum_{i=1}^n
(t_i^\ast)^{\widehat{\beta}}}{C}\biggr)^{{1}/{\widehat{\beta}}}.
\]
%
\subsection{\texorpdfstring{Derivation of the exact MLE for i.i.d. exponential case.}{Derivation of the exact MLE for i.i.d. exponential case}}
We assume $X,T\stackrel{\mathrm{i.i.d.}}{\sim}\operatorname{Exp}(\lambda)$. The complete
likelihood is given by
\[
L(\lambda) \propto(\lambda)^{2C} e^{-\lambda\sum_{i=1}^C
(x_i+ t_i)} [ e^{-\lambda T_0} + T_0 \lambda e^{-\lambda T_0}
]^{N-C}.
\]
Now differentiating the log-likelihood equation with respect to
$\lambda$ and equating it to zero yields
\[
\frac{2C}{\lambda}+\frac{T_0(N-C)}{1+\lambda T_0}-\sum_{i=1}^C (x_i
+ t_i)-(N-C)T_0=0.
\]
The equation needs to be solved numerically for $\lambda$ to obtain
MLE.
\end{appendix}
\section*{Acknowledgments}

Special thanks to Dr. Eric Adams for proposing the problem and for
his many valuable comments. I would also like to thank an anonymous
referee and the Associate Editor, whose comments provided additional
insights and have greatly improved the scope and presentation of the
paper.

\begin{supplement} 
\stitle{Furnace Data Set and R Code for Furnace Data as well as
Simulation for all Models Considered in the Paper}
\slink[doi]{10.1214/10-AOAS348SUPP} 
\slink[url]{http://lib.stat.cmu.edu/aoas/348/supplement.zip}
\sdatatype{.zip}
\sdescription{
R~code is used for the simulation as well as real data analysis.\\
\indent Supplementary material has five files:\vspace*{6pt}\\
\indent 1. Furnace data in MS Excel format (data.xls).\\
\indent 2. Code for analyzing furnace data (code\_furn.doc).\\
\indent 3. Code for the Exponential--Exponential model
(new\_code\_Exp(2).doc).\\
\indent 4. Code for the Exponential--Weibull model
(new\_code\_ExpWeb.doc).\\
\indent 5. Code for the Weibull--Exponential model
(new\_code\_WebExp.doc).\vspace*{6pt}\\
For the simulation examples data sets are generated on the fly at the
beginning of the code. No special R package is required to run the
codes. All the codes are commented for the ease of understanding.}
\end{supplement}

\printaddresses

\end{document}